\documentclass[twocolumn]{aastex62}

\usepackage{epsfig} 
\usepackage{graphicx,times,subfigure}
\usepackage{amsmath}
\usepackage{natbib}
\usepackage{xfrac}
\usepackage[T1]{fontenc}
\usepackage{aecompl}
\usepackage{amssymb}
\usepackage{hyperref}

\def\apjl{ApJL}

\def\apjs{ApJS}  
\def\aap{A\&A}

\newcommand{\sigsfr}{\Sigma_\mathrm{SFR}}
\newcommand{\siggas}{\Sigma_\mathrm{gas}}

\newcommand{\tff}{t_\mathrm{ff}}

\newcommand{\gastmff}{(\siggas/t)_\mathrm{multi-ff}}
\newcommand{\mach}{\mathcal{M}}

\graphicspath{{./}{figures/}}

\begin{document}

\title{Spinning Bar and a Star-formation Inefficient Repertoire: \\ Turbulence in Hickson Compact Group NGC7674}

\correspondingauthor{Diane M. Salim}
\email{diane@mso.anu.edu.au}

\author{Diane M. Salim}
\affiliation{Research School of Astronomy and Astrophysics, Australian National University, Canberra, ACT 2611, Australia} 
\affiliation{ARC Centre Of Excellence For All Sky Astrophysics in 3D (ASTRO3D), Australia} 

\author{Katherine Alatalo}
\affiliation{Space Telescope Science Institute, 3700 San Martin Dr., Baltimore, MD 21218, USA}
\affiliation{The Observatories of the Carnegie Institution for Science, 813 Santa Barbara St., Pasadena, CA 91101, USA}
\affiliation{Johns Hopkins University, Department of Physics and Astronomy, Baltimore, MD 21218, USA}

\author{Christoph Federrath}
\affiliation{Research School of Astronomy and Astrophysics, Australian National University, Canberra, ACT 2611, Australia} 
\affiliation{ARC Centre Of Excellence For All Sky Astrophysics in 3D (ASTRO3D), Australia} 

\author{Brent Groves}
\affiliation{Research School of Astronomy and Astrophysics, Australian National University, Canberra, ACT 2611, Australia} 
\affiliation{ARC Centre Of Excellence For All Sky Astrophysics in 3D (ASTRO3D), Australia} 

\author{Lisa J. Kewley}
\affiliation{Research School of Astronomy and Astrophysics, Australian National University, Canberra, ACT 2611, Australia} 
\affiliation{ARC Centre Of Excellence For All Sky Astrophysics in 3D (ASTRO3D), Australia}

\begin{abstract}
The physics regulating star formation (SF) in Hickson Compact Groups (HCG) has thus far been difficult to describe, due to their unique kinematic properties. In this study we expand upon previous works to devise a more physically meaningful SF relation able to better encompass the physics of these unique systems. We combine CO(1--0) data from the Combined Array from Research in Millimeter Astronomy (CARMA) to trace the column density of molecular gas $\siggas$ and deep H$\alpha$ imaging taken on the Southern Astrophysical Research (SOAR) Telescope tracing $\sigsfr$ to investigate star formation efficiency across face-on HCG, NGC7674. We find a lack of universality in star formation, with two distinct sequences present in the $\siggas-\sigsfr$ plane; one for inside and one for outside the nucleus. We devise a SF relation based on the multi-freefall nature of gas and the critical density, which itself is dependent on the virial parameter $\alpha_{\mathrm{vir}}$, the ratio of turbulent to gravitational energy. We find that our modified SF relation fits the data and describes the physics of this system well with the introduction of a virial parameter of about 5--10 across the galaxy. This  $\alpha_{\mathrm{vir}}$ leads to an order-of-magnitude reduction in SFR compared to $\alpha_{\mathrm{vir}}\approx 1$ systems. 
\end{abstract}


\section{Introduction}\label{sec:intro}

Star formation is complicated. The entangled, mutually dependent relationship between gas and star formation is complex and non-linear. Locally, the coalescence of gas via the influence of turbulence and gravity serve to regulate star formation within the coldest phase of the interstellar medium (ISM) \citep{Ferriere:2001aa,MacLowKlessen2004,ElmegreenScalo2004,McKeeOstriker2007,PadoanEtAl2014}. Whilst the underlying physics that determine the column density of star formation ($\sigsfr$) is still not known exactly and has been under heated debate for more than two decades, observations have shown that SF regulation follows patterns. Previous relations have suggested parameterizations of SF including the column density of available gas ($\siggas$) \citep{Schmidt1959, Kennicutt:1998aa, Bigiel:2008aa, LeroyEtAl2008,DaddiEtAl2010a, SchrubaEtAl2011, RenaudEtAl2012, Kennicutt:2012aa}, such as the \citet{Kennicutt:1998aa} (hereafter K98) SF power-law relation: 
\begin{align}
\sigsfr \propto \siggas^{1.4}~
\end{align}
 and the \citet{KrumholzDekelMcKee2012} (hereafter KDM12) relation, which correlate $\sigsfr$ to the ratio between the mean column gas density and the mean free-fall time $\tff(\rho_0)$ of the cloud, which is the time scale required for a medium with negligible pressure support to collapse. By doing so, the self-gravity of the system is taken into account: 
\begin{align}
\sigsfr =  \epsilon_{\mathrm{ff}}\frac{\siggas(\rho_0)}{\tff(\rho_0)} ~,
\end{align}
where $\epsilon_{\mathrm{ff}}$ is the efficiency of the gas. Calibrating for Milky Way (MW) clouds and Local Group galaxies, this efficiency is $\approx 1\%$. 

Many studies have also proposed and observed supersonic random motions to play a key factor in regulating star formation \citep{ZuckermanEvans1974, ZuckermanPalmer1974, Larson1981, SolomonEtAl1987, FalgaronePugetPerault1992, OssenkopfMacLow2002, HeyerBrunt2004, SchneiderEtAl2011, RomanDuvalEtAl2011}. The two-fold role of turbulence can be seemingly contradictory. Turbulence may serve as a SF suppressor, because turbulent kinetic energy stabilizes molecular clouds on large scales to prevent global collapse. However, these SF suppression effects are juxtaposed against the potential for turbulence to enhance SF. Turbulence is supersonic, thus may induce local compressions in shocks \citep{MacLowKlessen2004, ElmegreenScalo2004, McKeeOstriker2007}. This in turn produces filaments and therefore dense cores at the intersections of filaments, thereby generating the initial conditions for SF \citep{SchneiderEtAl2012}. 

There have been extensive studies investigating relations between turbulence, molecular gas and star formation in Milky Way clouds \citep{HeidermanEtAl2010,LadaEtAl2008,WuEtAl2010,Bigiel:2008aa,Bigiel:2011aa}. The influence of turbulence on the SF process is most apparent in the observations of giant molecular clouds (GMCs) in the MW, which are widely accepted to be supersonically turbulent structures \citep{Larson1981, ElmegreenScalo2004, MacLowKlessen2004, McKeeOstriker2007, HennebelleFalgarone2012, PadoanEtAl2014}. GMCs do not collapse freely to form stars themselves, because the rate of gas conversion over a cloud's freefall time would result in a star formation rate two orders of magnitude more than that observed in the MW \citep{ZuckermanEvans1974,KrumholzTan2007}. Whilst this could be due to stellar feedback, simulations have failed to produce this scenario \citep{HopkinsEtAl2012,Tasker:2015aa,Howard:2016aa}. The other viable explanation is that the additional energy from internal turbulence or magnetic fields regulate cloud dynamics to hinder the rapid conversion of gas into stars \citep{FederrathKlessen2012,PadoanEtAl2014}. The force required to drive the turbulence can originate from both within the cloud via stellar feedback and outside the cloud from shear or interactions with neighbouring clouds \citep{Federrath2016CMZbrick, Federrath2017IAUsymposium}. Theories of star formation that incorporate turbulence have also yielded good matches to observations for predictions of the initial mass function of stars \citep{PadoanNordlund2002,HennebelleChabrier2011,HennebelleChabrier2013} as well as star formation rates and efficiencies \citep{KrumholzMcKee2005, PadoanNordlund2011, HennebelleChabrier2011, FederrathKlessen2012, FederrathKlessen2013, Federrath2013, PadoanEtAl2014, SalimEtAl2015, Sharda:2018aa}. This in turn serves as an indication that GMC evolution is dictated by gravity and turbulence, influenced by the environment of their host galaxy. Furthermore, 
there have been extensive efforts to elucidate what makes the Central Molecular Zone (CMZ), the central 500 pc of the MW, so inefficient in forming stars. Despite high gas densities and large gas reservoirs, the CMZ exhibits SF which is about an order of magnitude less active than expected \citep{LongmoreEtAl2013, JohnstoneEtAl2000, Kruijssen:2014aa}, and many studies have suggested turbulence to play a central role in this supression \citep{RathborneEtAl2011, RathborneEtAl2014, Kruijssen:2015aa, Federrath2016CMZbrick, Sormani:2018aa}. Studying the effects of turbulence is therefore integral to understand galaxies' evolutionary histories.

Unfortunately it is difficult to resolve turbulence on extragalactic scales and especially more so at high redshift. To properly investigate the effects of turbulence one would require resolution to at least the cloud scale of order 10 pc. Whilst studies of such scales are possible to investigate regions of the Milky Way \citep{Federrath2016CMZbrick, RathborneEtAl2011, RathborneEtAl2014} and in one exceptional case so far also for a high-redshift lensed galaxy \citep{Sharda:2018aa}, this is beyond the reach for many galaxies. Studies of turbulence and SF in extragalactic systems therefore necessitate studying objects with highly enhanced turbulent properties to maximise signal to noise in order to compensate for lower spatial resolutions.  Many galaxies are found to exist within larger structures such as groups or clusters, the members of which also significantly impact neighbouring galaxies' intrinsic properties, thus serving as a great case study to investigate systems whose internal turbulence is affected by a variety of physical processes. 

To directly probe the enhanced turbulent features in unique grouped galaxy environments we explore the kinematics of a Hickson Compact Group (HCG) galaxy. Grouped galaxies like HCGs make for ideal case studies to investigate the interactions between vigorous galaxy transition, SF quenching and turbulence because we know that they are undergoing that change. HCGs are defined as small groups of several (4 or more) galaxies separated by projected distances of only a few tens of kiloparsecs (comparable to galaxy sizes) and relatively isolated from other large structures \citep{Hickson:1982aa,Hickson:1997aa,Bitsakis:2014aa}. HCGs are therefore some of the densest known galaxy systems. HCGs are similar in nature and structure to the central regions of very dense galaxy clusters \citep{Hickson:1982aa,Hickson:1997aa},  exhibiting high densities but low intergalactic velocity dispersions \citep{Hickson:1997aa}. HCGs appear to transform very rapidly from star-forming to quiescent \citep{Johnson:2007aa, Walker:2010aa}. 

HCGs have also been shown to exhibit star formation suppression. \citet{AlataloHCG2015} (hereafter A15) used the Combined Array from Research in Millimeter Astronomy (CARMA) to observe and map CO(1-0) of galaxies belonging to 12 HCGs. The sample studied was chosen to follow up a subset studied in \citet{Lisenfeld:2014aa} of CO-bright, warm $\mathrm{H}_2$-bright HCG galaxies. The authors found that the global efficiencies of a large portion of these galaxies are up to two orders of magnitude less than that predicted by the K98 SF power-law and KDM12 single-freefall model for star formation.

In addition, HCG galaxies have been shown to exhibit signs of enhanced turbulence via strong shocks. \citet{Cluver:2013aa}'s study using the \textit{Spitzer} Infrared Spectograph find that their sample of HCG galaxies tend to exhibit warm hydrogen emission enhanced star-formation-powered photon-dominated regions, due to shocks caused by collisions with the clumpier inter-group medium. By using  data from \citet{Cluver:2013aa}, A15 showed that the galaxies' warm $\mathrm{H}_2$ luminosities are within a factor 3 of the turbulent injected energy. The observed shocks and high levels of turbulence may share the same origin, both being driven by the galaxy interactions. This additional turbulence could create an energy imbalance which leads to star formation suppression. Similar systems with excess turbulence and inefficient star formation have been observed in the Milky Way \citep{KauffmannEtAl2013}, NGC1266 \citep{Alatalo:2015ab}, and 3C 326N \citep{Guillard:2015aa}. 

We study a face-on HCG from the sample in A15 with an inclination of 26.7 $\deg$, namely NGC7674. In this work we isolate a single HCG from A15 to investigate turbulent motions within a galaxy using the spatially resolved CARMA CO(1-0) flux maps from A15 and deep H$\alpha$ imaging from the SOAR Telescope presented by \citet{Eigenthaler:2015aa}.  The molecular gas in NGC7674 is morphologically classified as a spiral and bar+ring, due to its multiple components. It contains an AGN, classified from its mid-IR spectrum \citep{Cluver:2013aa} and optical spectroscopy \citep{Martinez:2010aa}. NGC7674 is interacting with its HCG companion HCG96c, exhibiting numerous tidal tails and stellar light bridging the two systems. 

In Section~\ref{sec:obs} we describe our data and its preparations for analysis. In Section~\ref{sec:results} we describe the direct results of the observed data and the relationship between $\siggas$ and $\sigsfr$ in this galaxy. In Section~\ref{sec:veldisp} we describe our method of isolating the turbulent motions of the galaxies and present these results. In Section~\ref{sec:sfr_laws} we use the inferred $\siggas$ data to predict $\sigsfr$ using various SF relations, and compare them to the observed $\sigsfr$. In Section~\ref{sec:discussion} we explore the implications of the SF models which best described the physics of NGC7674. Section~\ref{sec:conc} we summarise our results.

Throughout this study we assume the \citet{PlanckCollaboration2014} Hubble constant of 67.80 $\mathrm{km~s^{-1} Mpc^{-1}}$. In all the spatial images presented in this work North is up and East is left. 

\begin{figure*}
\includegraphics[width=\linewidth]{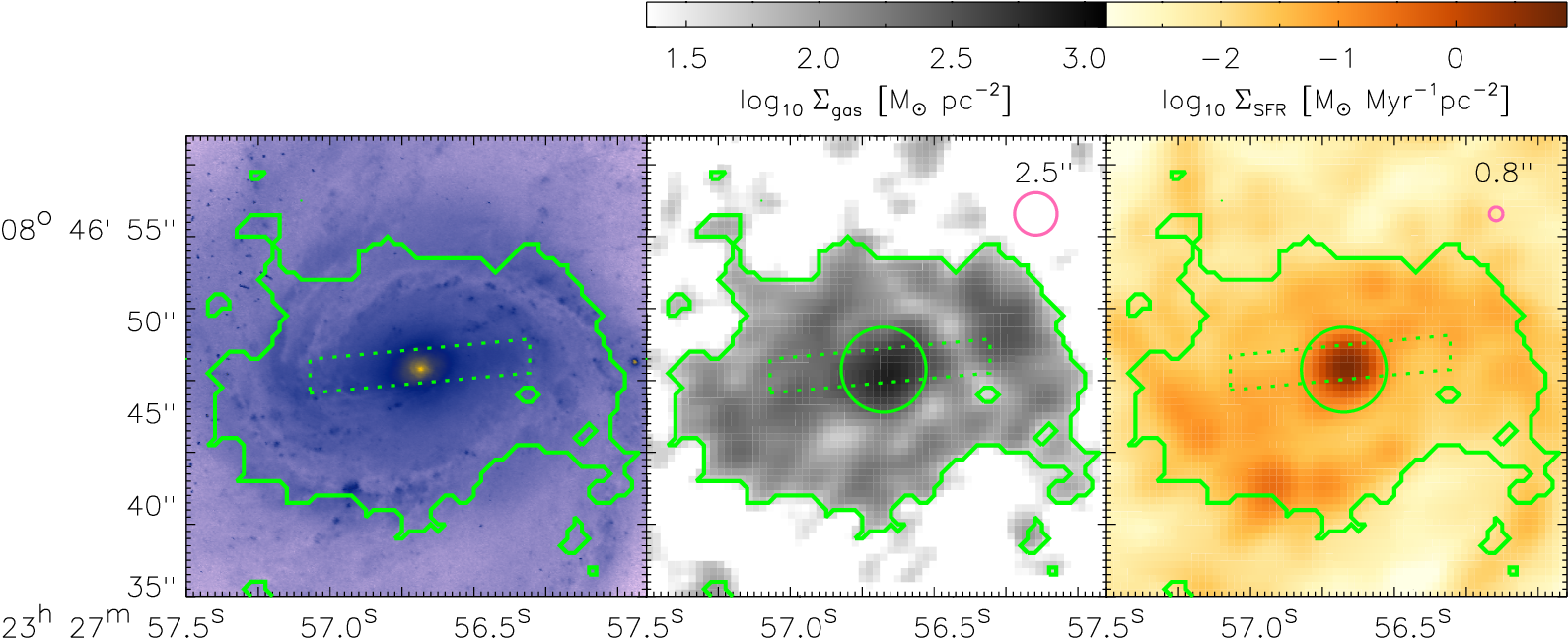} 
\caption{Hickson Compact Group NGC7674. \emph{LEFT:} $i$-band image from the \textit{Hubble Space Telescope (HST)} \citep{Armus:2009aa}. \emph{MIDDLE:} $\siggas$ maps inferred from the CARMA CO(1-0) observations \citep{AlataloHCG2015}. \emph{RIGHT:} $\sigsfr$ maps inferred from deep H$\alpha$ imaging taken with the 4.1m SOAR telescope. \citep{Eigenthaler:2015aa}. In every panel, the outermost contour indicates the boundary at which we seperate the noise-dominated pixels to the signal-dominated pixels we use in analysis. The dotted green rectangle indicates the boundary of the bar of NGC7674, as visually determined by the HST $i$-band image. In the middle and right panels, the central green circle represents the boundary of the central nucleus, as determined by the deep H$\alpha$ imaging. In the middle and right panels, the top-right pink circle indicates the FWHM of the beam of the instruments which took the observations. The FHWM of each respective instrument is written above the beam size.}
\label{fig:maps}
\end{figure*}

\section{Observations and data preparation} \label{sec:obs}
\subsection{Data Acquisition}
 

To thoroughly probe the interplay between molecular gas, turbulence and SF in NGC7674 on small scales, we need spatially resolved images of both $\siggas$ and $\sigsfr$, as well as information regarding gas kinematics. For this reason we choose to utilise interferometric CO(1--0) data taken with CARMA\footnote{\href{http://www.mmarray.org}{http://www.mmarray.org}} from A15 and partner this with deep H$\alpha$ imaging taken with the 4.1m Southern Astrophysics Research (SOAR) Telescope form \citet{Eigenthaler:2015aa}. 

CARMA was an interferometric array of 15 radio dishes each of dimensions of 6$\times$10.4m or 9$\times$6.1m located in the Eastern Sierras in California \citep{BockEtAl2006}. The galaxy was observed using CARMA's D-array with baselines between 11 to 50 m. \citet{AlataloEtAl2013} describes in detail the procedure to observe and reduce the data for NGC7674. This method resulted in the spatially resolved moment and channel maps analysed in this study. We use the 0th moment map of this galaxy to infer the distribution of $\siggas$ across the galaxy, and we use the channel map and the velocity spectrum within each pixel to infer the kinematic properties of the galaxy. These maps have a full width half-maximum (FWHM) of 2.5 arcseconds and a spatial scale of 0.4 arcseconds per pixel.

The SOAR Telescope is a 4.1 meter aperture telescope located on Cerro Pach\'{o}n in Chile. The Goodman Spectograph was used in imaging mode to measure H$\alpha$ + [NII] narrowband fluxes, with a spatial scale of 0.15 arcseconds per pixel, with an average seeing of 0.8 arcsec during the night of observation. Full details of the data acquisition and reduction techniques used for these data are available in \citet{Eigenthaler:2015aa}

\subsection{Inferring $\siggas$} \label{sec:siggas}
We estimate the column density of molecular hydrogen $\mathrm{H_2}$, traced by the carbon monoxide CO(1--0) transition line observed by CARMA. We determine the column density of molecular hydrogen in each line of sight (LOS) by:
\begin{equation}
M(\mathrm{total ~H_2 ~in~LOS}) ~[\mathrm{g~cm^{-2}}] = S_{\mathrm{CO}}\Delta v \cdot K \cdot X_{\mathrm{CO}} \cdot M_{\mathrm{H_2}}, 
\end{equation}
where $S_{\mathrm{CO}}\Delta v$ is the CO(1-0) flux in $\mathrm{Jy~km~s^{-1}}$, $K$ is the Kelvin per Jansky factor of 15.02 as quoted in \citet{AlataloHCG2015}. The assumed $X_{\mathrm{CO}}$ CO-to-$\mathrm{H_2}$ factor is $2\times10^{20} ~\mathrm{cm^{-2}(K~km^{-1})^{-1}}$, which is the value presented in \citet{BolattoEtAl2013}. The mass of a single molecular hydrogen atom $ M_{\mathrm{H_2}}$ is $1.66\times10^{-24}$ g. Whilst we acknowledge that the centres of galaxies with strong bars tend to exhibit a smaller gas-to-dust conversion factor than that in the disk \citep{Sandstrom:2013aa, BolattoEtAl2013}, \citet{AlataloHCG2015} showed that the gas-to-dust ratio of NGC7674 fell within the range for nearby galaxies of solar metallicities and exhibited little to no dependence on star formation suppression. This indicates that it is acceptable to apply a standard Galactic conversion factor for inferring the $\siggas$ content. Detailed observations of $^{12}$CO isotopologues such as $^{13}$CO and C$^{18}$O and dense gas such as HCN, HCO$^{+}$ and CS are required to confirm whether the system requires a different conversion factor.

We obtain our final $\siggas$ values by converting the total $\mathrm{H_2}$ mass in each pixel into units of $\mathrm{M_{\odot}~pc^{-2}}$ then multiplying by 1.36, which incorporates the correction factor to account for Helium. The final spatial distribution of $\siggas$ across the galaxy is shown in the middle panel of Figure~\ref{fig:maps}. 

\subsection{Inferring $\sigsfr$} \label{sec:sigsfr}
The column density of SF was traced by H$\alpha$+[NII] fluxes measured by the SOAR telescope. Following \citet{Eigenthaler:2015aa}, we apply two corrections to better isolate the flux specifically originating from H$\alpha$. Firstly, the narrowband fluxes must be corrected for the contamination of the nitrogen [NII]$\lambda\lambda 6548,6583$ lines. A H$\alpha$/(H$\alpha$+[NII]) ratio of 0.74$\pm$0.13 was applied, which is based on integrated [NII]/H$\alpha$ measurements of 58 galaxies from the SINGS survey \citep{Kennicutt:2003aa, Kennicutt:2009aa}. 

Secondly, we correct for dust extinction of H$\alpha$ flux by taking advantage of the relation between H$\alpha$ extinction and stellar mass \citep{GarnBest2010,Kashino:2013aa}. We use the \citet{GarnBest2010} relation that estimates the extinction in H$\alpha$, $A_{\mathrm{H\alpha}}$ as:
\begin{align}
A_{\mathrm{H\alpha}} = 0.91 + 0.77X + 0.11X^2 - 0.09X^3,
\end{align} 
where $X = \log_{10}(\mathrm{M_*/M_{\odot}}) - 10$ with an uncertainty quoted as 0.28 mag. \citet{Eigenthaler:2015aa} reports NGC7674 to have a stellar mass of $\log_{10}(\mathrm{M_*/M_{\odot}}) = 11.34\pm0.14$. The H$\alpha$ flux was then corrected for extinction using the \citet{CardelliClaytonMathis1989} reddening curve. Due to the known presence of H$\alpha$-enhancing shocks in this galaxy, the H$\alpha$ measurements are an upper limit. However, we confirm that this method adequately corrects for the dust extinction of the galaxy by comparing the final total extinction-corrected star formation rates to that obtained by tracing $\sigsfr$ by 70 $\mu$m emission as taken by the Herschel Photodetecting Array Camera and Spectrometer (PACS) instrument and first presented in \citet{Bitsakis:2014aa}. 70 $\mu\mathrm{m}$ emission can be a good tracer for SF because it does not suffer from extinction. However, due to the PACS instrument's larger beamsize of 5.8" (compared to the SOAR Goodman Spectrograph's 0.8" and CARMA'S 2.5"), the 70$\mu\mathrm{m}$-inferred $\sigsfr$ map is incapable of resolving individual HII regions so a lot of information regarding the internal SF structure within NGC7674 would not be inferrable and thus inappropriate for our pixel-by-pixel study. We thus use the global values of 70$\mu\mathrm{m}$-inferred $\sigsfr$ to ensure that the extinction correction applied to the H$\alpha$ imaging-inferred $\sigsfr$ is adept enough to report an accurate measurement of $\sigsfr$. \citet{Bitsakis:2014aa} reports a total SFR of 17.4 M$_{\odot}$yr$^{-1}$ for NGC7674, whilst the total SFR in the H$\alpha$ imaging-inferred $\sigsfr$ map within the contours of signal-dominated pixels is is 15.7 M$_{\odot}$yr$^{-1}$. These total SFRs derived using these two methods agree within $\approx10\%$. We can therefore trust that the extinction correction applied adequately corrects for dust attenuation in NGC7674 such that we can trust the $\sigsfr$ values used in the subsequent analysis of this study.

Having implemented these corrections, the final corrected H$\alpha$ flux ($F_{\mathrm{H\alpha}}$) in each line of sight was converted to a luminosity through the relation: 
\begin{align}
L_{\mathrm{H\alpha}} ~[\mathrm{erg~s^{-1}}]= 4D_L^2 \pi F_{\mathrm{H\alpha}},
\end{align}
where $D_L$ is the luminosity distance of NGC7674 in cm. This luminosity was converted into a SFR following the calibration presented in \citet{Calzetti:2007aa} and reviewed in \citet{Kennicutt:2012aa}:
\begin{align}
\mathrm{SFR} ~[\mathrm{M_{\odot}~yr^{-2}}] = 5.3\times10^{-42} L_{\mathrm{H\alpha}}~[\mathrm{erg~s^{-1}}]
\end{align}
We use the IDL routine \texttt{HASTROM}\footnote{\href{https://idlastro.gsfc.nasa.gov/ftp/pro/astrom/hastrom.pro}{https://idlastro.gsfc.nasa.gov/ftp/pro/astrom/hastrom.pro}} to match the pixel scale of the SFR map to that of the CARMA pixel grid. This routine applies a transformation to an image so that its astrometry is identical with that in a reference header. We then convert our SFR into a column density of SF, $\sigsfr$ by dividing SFR by the galactocentric scale of $g$ of 0.586 kpc arcsec$^{-1}$, as recorded on the NASA/IPAC Extragalactic Database \footnote{\href{http://ned.ipac.caltech.edu/}{http://ned.ipac.caltech.edu/}}, which yields: 
\begin{align}
\sigsfr ~[\mathrm{M_{\odot}~yr^{-2}~kpc^{-1}}]= \frac{\mathrm{SFR}}{g \cdot w},
\end{align}
where $w$ is the pixel width of the CARMA grid of 0.4 arcseconds. The final spatial distribution of $\sigsfr$ across the galaxy is shown in the middle panel of Figure~\ref{fig:maps}. Using this image we visually determine the boundaries of the nucleus. The nucleus is demarcated as a circle in the middle and right panels of Figure~\ref{fig:maps}. 

\subsubsection{Influence of the AGN on H$\alpha$ as a star formation rate tracer}
\label{sec:AGNinfluence}
We acknowledge the undeniable presence of an AGN in the centre of NGC7674 that has been proven through infra-red \citet{Cluver:2013aa} and optical \citet{Martinez:2010aa} observations. H$\alpha$ emission is known to be enhanced by the influence of AGNs \citep{KewleyEtAl2002, Kewley2006aa, Kewley:2003aa, Rich:2011aa} so have traditionally been treated with caution as a tracer for SF in AGN-driven galaxies. This is especially an issue when the beam size of the observation instrument covers a greater physical area than that of the central nucleus of the galaxy, because the central concentration of H$\alpha$ emission (and thus inferred SFR) will be spread out via beam smearing and thus incorrectly attributed to SF in the disk.

However, given that NGC7674 is a Seyfert galaxy exhibiting particularly strong star formation activity \cite{Eigenthaler:2015aa} we can infer that the AGN emission is highly obscured by dust clouds forming stars. Furthermore, from the HST $i$-band image depicted in Figure~\ref{fig:maps}, it is evident that the very central nucleus exhibits an independent spiral structure finer than the resolution of the H$\alpha$ map. Therefore, in H$\alpha$, the central AGN can be treated as a point source the size of the beam of the instrument used to capture the H$\alpha$ emission. Thus to mitigate AGN influence on our star formation study, we have disregarded the pixels that fall within the area of the central beam in H$\alpha$ from further analysis.

\subsection{Phenomenological features and signal to noise cuts} \label{sec:noise}
We define the noise-dominated pixels as those that have a signal to noise ratio of less than 5 in the intensity-weighted mean velocity of CO along the line of sight (moment 1) observations. We define the lower threshold in $\siggas$ and $\sigsfr$ as the standard deviation of the noise-dominated $\siggas$ and $\sigsfr$ pixels respectively. These thresholds are shown in the KS plane as the vertical and horizontal dotted purple relations in Figure~\ref{fig:KScontours}. Any pixels that fall below these thresholds in either $\siggas$ or $\sigsfr$ are disregarded from subsequent analyses in the rest of this study. Thus to be considered signal-dominated and therefore adept for analysis a pixel must exhibit higher values of $\siggas$ \textit{and} $\sigsfr$ than the lower thresholds defined in both domains. The boundary between noise-dominated and signal-dominated pixels are shown spatially as the thickest, outer-most contour in every panel of Figure~\ref{fig:maps}. Pixels that fall within this contour are considered signal-dominated in both $\siggas$ and $\sigsfr$, whilst anything outside is considered noise-dominated and thus omitted form analysis. Any subsequent figures following Figure~\ref{fig:maps} showing the spatial distribution of $\siggas$ across NGC7674 will only show the data which falls within the boundary of lines of sights that are not noise-dominated. 

In order to investigate whether these regions have properties distinctly differing from the rest of the galaxy, we isolate the regions in which the galactic bar and nucleus lie. In the left-most panel of Figure~\ref{fig:maps} we show the optical $i$ band image from the \textit{Hubble Space Telescope}, observed as part of the Great Observatories
All-sky LIRG Survey (GOALS) \citep{Armus:2009aa}. Using this image we visually determine the boundaries of the bar. A rectangle denotes the bar in each panel of Figure~\ref{fig:maps} and every subsequent image of NGC7674 in this study. This galaxy is also predicted to host compact nuclear gas. We visually determine the boundary of the nucleus via the final $\sigsfr$ map. In each panel of Figure~\ref{fig:maps} and every subsequent image of NGC7674 in this study the nucleus is demarcated by a circle. We show the entire bar for visualisation purposes, but for subsequent analysis pixels of the bar which fall within the boundary of the nucleus are classified as only a nucleus pixel. 


\section{Results: interaction between gas and star formation} \label{sec:results}
\subsection{Behaviour in the KS plane}

\begin{figure*}
\includegraphics[width=\linewidth]{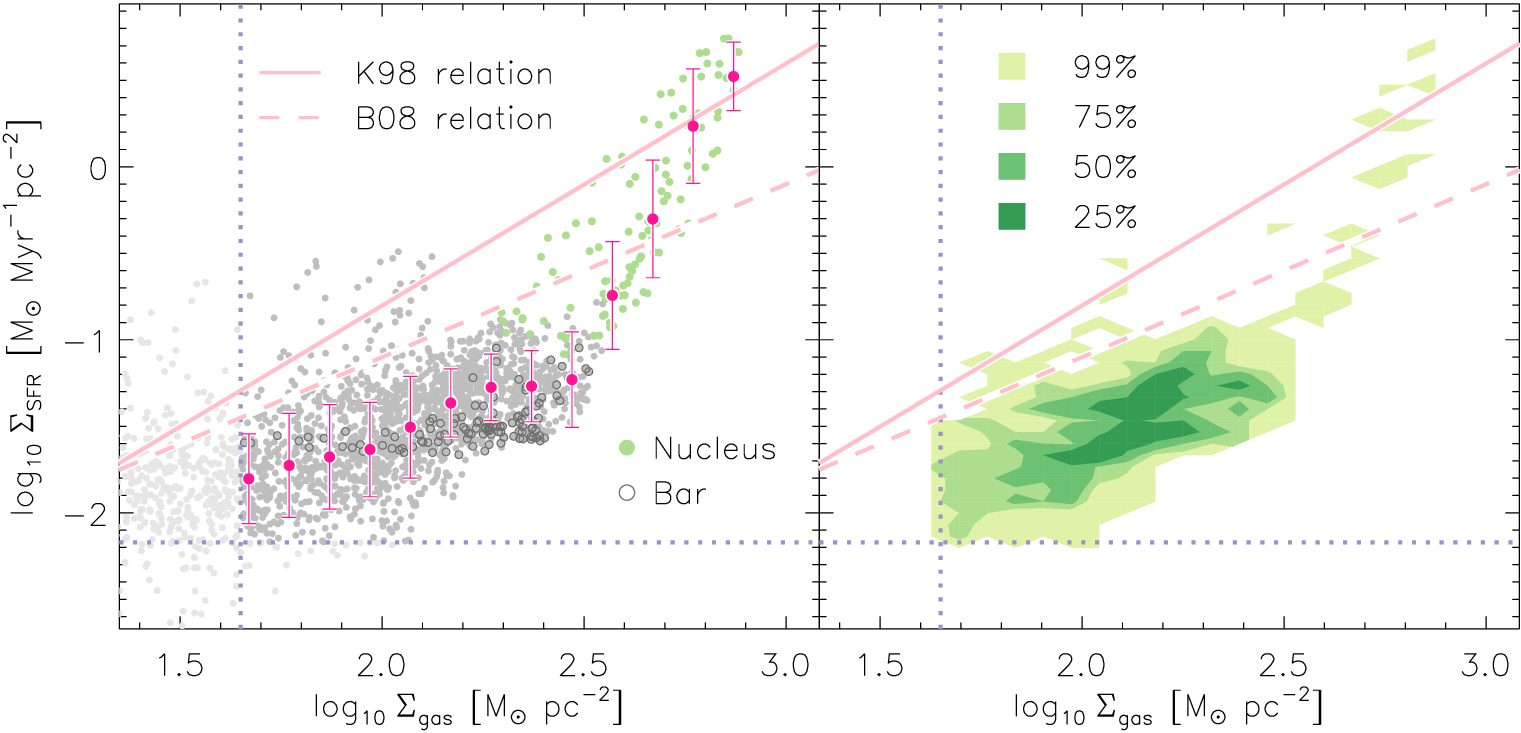}
\caption{\textit{LEFT:} We show the molecular gas surface density from CO(1--0) of every line of sight column in NGC7674's CARMA field compared to the $\sigsfr$ for the same column. Each LOS column is represented as a single circular data point in the left panel of the diagram. Green circles indicate pixels that fall within the bounds of the galactic nucleus and outlined circles highlight pixels that make up the bar outside the nucleus (see Fig.~\ref{fig:maps} for the parts of the galaxy defined as nucleus and bar, respectively). Both of these regions are defined in Section~\ref{sec:noise}. The overlaid dark pink circles are the median $\sigsfr$ within $\siggas$ bins of 0.1 dex of $\mathrm{M_{\odot}pc^{-1}}$. The vertical error bars extending above and below these points are the standard deviations of $\sigsfr$ within each $\siggas$ bin. The \citet{Kennicutt:1998aa} (K98) and \citet{Bigiel:2008aa} (B08 hereafter) relations are also displayed on the same axes as the solid and dashed pink lines respectively. The lower thresholds in $\siggas$ and $\sigsfr$, as described in Section~\ref{sec:noise}, are also shown in Figure~\ref{fig:KScontours} as the dotted purple vertical and horizontal dashed lines respectively. Any values in either $\siggas$ or $\sigsfr$ that are lower than these thresholds we disregard and show in a fainter gray. Note that because the pixel size of the CARMA grid (0.4") is smaller than that of the PSF (2.5"), adjacent pixels are correlated. \textit{RIGHT:} We show the contours that represent the two-dimensional probability density within the Kennicutt-Schmidt framework of the same data as shown in the left-hand panel. Darker green areas therefore show regions of the KS plane that the LOS columns of NGC7674 are more likely to occupy. } 
\label{fig:KScontours}
\end{figure*}

While A15 demonstrate that this galaxy does not globally deviate significantly from the K98 relation, we find that on resolved scales, the bulk of NCG7674 is very inefficient at forming stars, as highlighted in Figure~\ref{fig:KScontours}. Because the global value of CO of a galaxy is a weighted average by luminosity, more luminous areas such as the galaxy's central region contribute a more significant influence in calculating the galaxy's global $\siggas$ and $\sigsfr$ properties. We therefore infer that the reason that the global value of NGC7674 lies so close to the K98 relation is because most of the lines of sight around the central region lie close to the K98 relation. This is apparent when we consider the nuclear region (green data points) labelled lines of sights in Figure~\ref{fig:KScontours}.

In this work we take advantage of the spatially resolved measurement of $\siggas$ and $\sigsfr$ to identify and analyze regions of inefficient SF. For any $\siggas$, the corresponding $\sigsfr$ can vary by more than an order of magnitude. Furthermore, it is apparent that two distinct SFR sequences exist which intersect at about $10^{2.3}~\mathrm{M_{\odot}pc^{-2}}$. The nuclear region (green data points) in the left panel of Figure~\ref{fig:KScontours} exhibits a much steeper $\siggas-\sigsfr$ slope than the bulk of the galaxy.
However, the low-$\siggas$ region, which contain the outskirts of the galaxy, is inefficient. We are confident in the physicality of these two distinct sequences because the beam size of the H$\alpha$ observation instrument is much smaller in area than the nuclear region defined in Section~\ref{sec:sigsfr}, and we have disregarded any pixels falling within the central beam in H$\alpha$, as described in Section~\ref{sec:AGNinfluence}. Therefore the steep SF relation within the nuclear region cannot merely be the systematic effect of beam smearing due to H$\alpha$-enhanced AGN emission.

The right-hand panel of Figure~\ref{fig:KScontours} illustrates that almost all lines of sight in this galaxy lie within the low-$\siggas$ region. In this region, the median $\sigsfr$ in each $\siggas$ bin consistently lies about 0.3-0.8 dex and 0.1-0.3 dex below the K98 and B08 SF relations, respectively. 

NGC7674 is a barred galaxy. We see that the pixels which make up NGC7674's bar outside of the nucleus follow an especially star-formation inefficient sequence. As a population, barred galaxies have been found to exhibit higher gas concentrations \citep{Sakamoto:1999aa, Jogee:2005aa, Sheth:2005aa, Regan:2006aa} and higher central SFRs \citep{de-Jong:1984aa, Hawarden:1986aa, Devereux:1987aa, Puxley:1988aa, Ho:1997aa} than most galaxies. Recent studies such as \citet{Chown:2018aa} and \citet{Lin:2017aa} find evidence that the presence of bar or tidal interaction is a necessary condition for central SF enhancement. Therefore, NGC7674's bar is likely to be driving gas from the outskirts of the galaxy into the nucleus. This would serve as a reasonable explanation for our finding that the centre of the galaxy exhibits a vastly different SF behaviour than the areas of the outskirts. This result exemplifies the limitations of inferring SF properties of galaxies from a single global data point (as in K98) across the galaxy. We need spatially resolved studies to investigate galaxy characteristics capable of reducing $\sigsfr$ in the bulk of the galaxy. 


\subsection{Spatial distribution of depletion times}

\begin{figure}
\includegraphics[width=\linewidth]{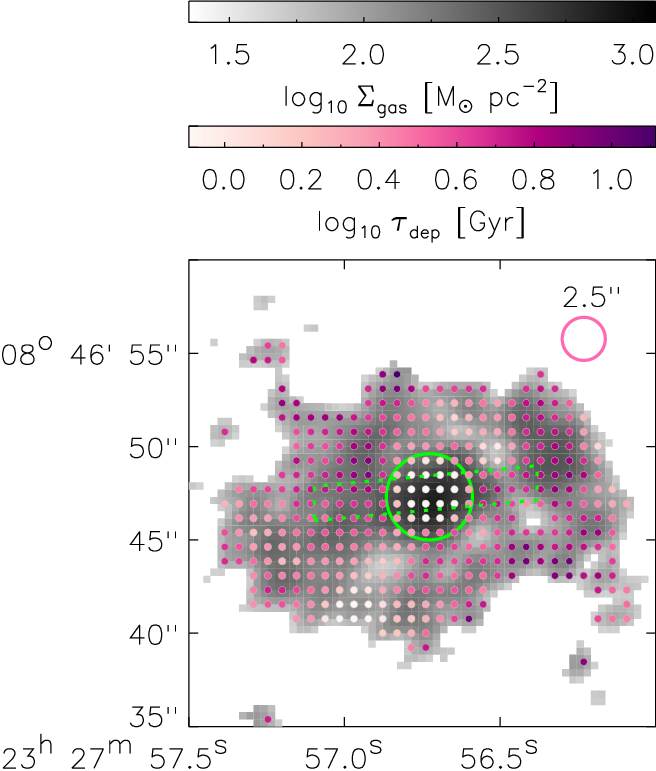} 
\caption{Average depletion time (pink colour gradient) of every $2\times2$ line of sight columns overlaid on top of the corresponding $\siggas$ map (grayscale). Only pixels above the $\siggas$ and $\sigsfr$ noise thresholds are shown.}
\label{fig:dots}
\end{figure}

NGC7674's lack of a universal $\siggas-\sigsfr$ relation is further emphasized by the spatial distribution of depletion time, $\tau_{\mathrm{dep}}$, across the galaxy. The depletion time is the rate at which GMCs form stars and is defined as the ratio between the molecular gas surface density and SF rate: 
\begin{align}
\tau_{\mathrm{dep}} = \frac{\siggas}{\sigsfr} .
\end{align}
This is the time that is required to turn the available gas into stars, also known as the timescale of SF. In Figure~\ref{fig:dots}, the grayscale map represents the $\siggas$ values of the pixel, which is also presented in Figure~\ref{fig:maps}. The overlaid pink dots indicate the average $\tau_{\mathrm{dep}}$ per $2\times 2$ pixels or 0.8$\times$0.8 arcsecs. Without external influences, one would expect that areas with the densest gas collapse the fastest and therefore have the shortest depletion times. Whilst this is true of the nucleus, there is no clear spatial correlation of these two factors in the rest of the galaxy. Aside from the central nuclear region, the densest gas is located in the outer regions of the galaxy. These areas, however, also host regions of some of the galaxy's longest depletion times of $>4$ Gyr. This further suggests that despite the abundance of dense gas, a competing factor must be suppressing SF. This supports many previous studies that have suggested that SF is not only a function of the molecular gas density. Other factors that have found to control SF rate include turbulence \citep{KrumholzMcKee2005, Federrath2010, PadoanNordlund2011, HennebelleChabrier2011, FederrathKlessen2012, Federrath2013sflaw, SalimEtAl2015}, magnetic fields \citep{Federrath:2015aa,Krumholz:2019aa} and free-fall time, which is the time scale required for a medium with negligible pressure support to collapse \citep{KrumholzDekelMcKee2012}. 

\section{Turbulent velocity dispersion} \label{sec:veldisp}


\begin{figure*}
\includegraphics[width=\linewidth]{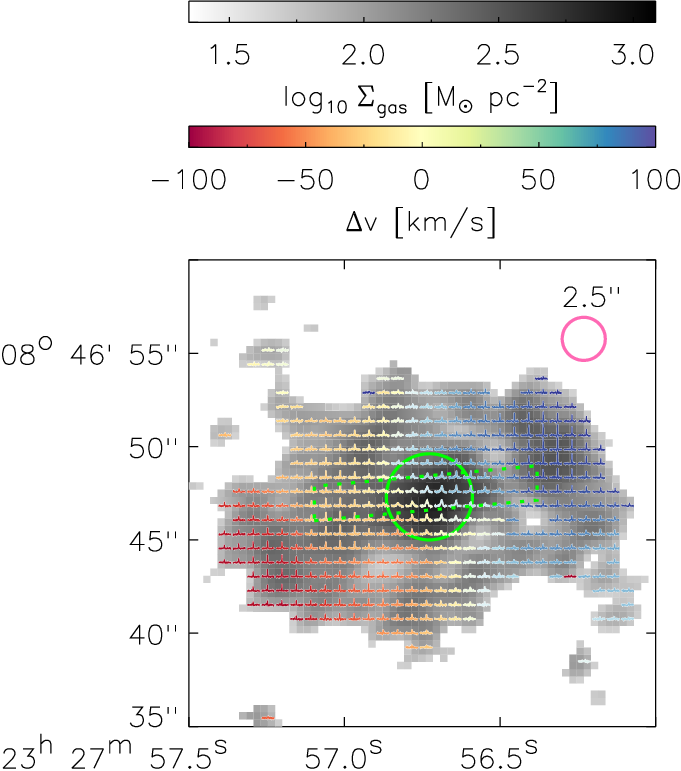}
\caption{Same as Figure~\ref{fig:dots}, but with integrated $2\times 2$ CO(1--0) spectra overlaid on the integrated intensity map (grayscale). Each spectrum was constructed by summing every pixel in the $2\times 2$ sub-grid in each channel. We fit each spectrum with Gaussian function as in Equation~\ref{eq:gaussian} and colour each spectrum from red to blue in accordance to the difference between the velocity of the Gaussian peak and the galaxy's average velocity, with red indicating redshifted and blue indicating blueshifted velocities.}
\label{fig:spectra}
\end{figure*}

\begin{figure*}
\includegraphics[width=\linewidth]{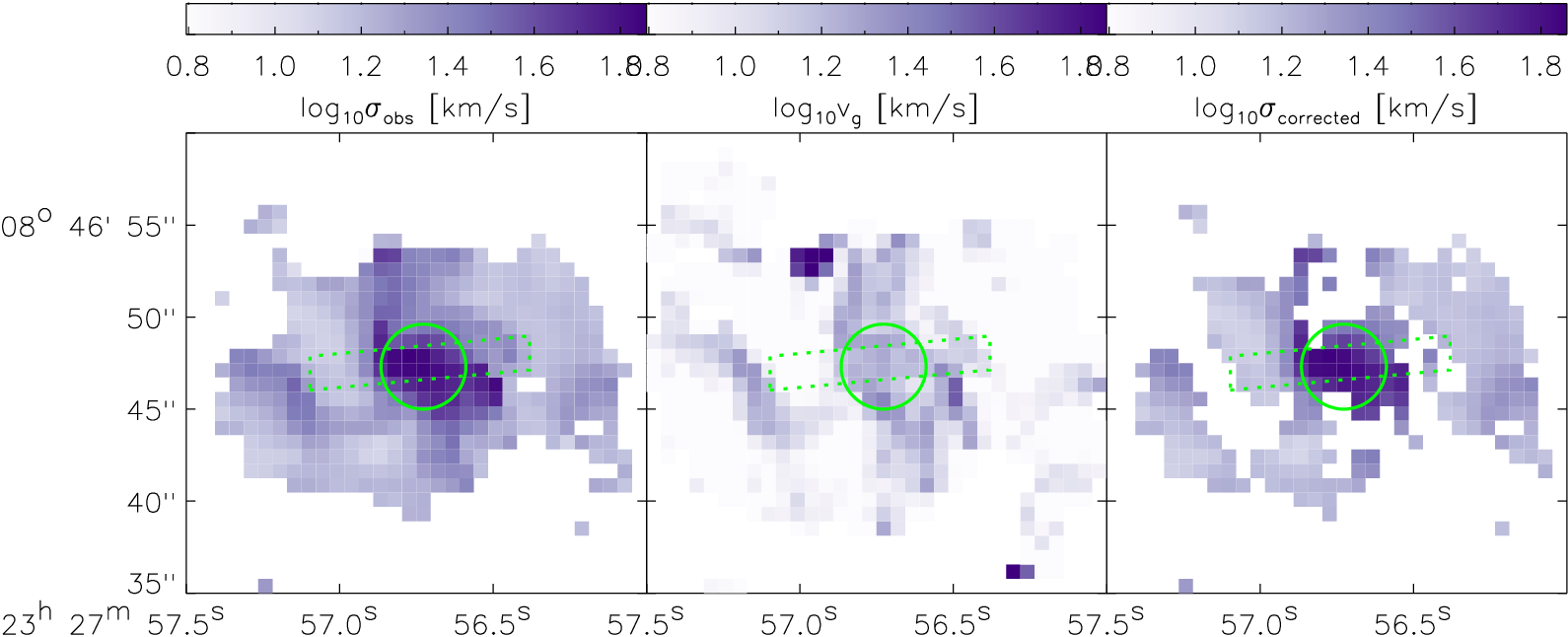} 
\caption{\emph{LEFT PANELS:} The uncorrected velocity dispersion (moment2) map, the values of which are derived from the standard deviation of the Gaussian for each spectrum shown in Figure~\ref{fig:spectra}. \emph{MIDDLE PANELS:} Velocity gradients across the galaxy. \emph{BOTTOM PANELS:} The gradient-corrected velocity dispersions of the galaxy (velocity gradient subtracted from fitted Gaussian standard deviations.)}
\label{fig:SD}
\end{figure*}

We investigate properties of the velocity distribution of the molecular gas to determine whether turbulent forces could play a significant role in inhibiting SF in this system.  We first measure the total velocity dispersion (2nd moment) across the galaxy using CARMA data. We increase signal-to-noise by creating spectra of $2\times 2$ binned pixels (rather than individual pixels). 

These spectra are shown in Figure~\ref{fig:spectra}, overlaid on the corresponding spatial area in $\siggas$. To each spectrum we fit a Gaussian $F$ as a function of velocity $v$:
\begin{equation}
F = A_0\exp\left[-\frac{1}{2}\left(\frac{v - A_1}{A_2}\right)^2 \right],
\end{equation} \label{eq:gaussian}
where $A_0$ is the height of the Gaussian peak, $A_1$ is the velocity at which the peak occurs and $A_2=\sigma$ is the standard deviation of the Gaussian distribution. Where there is a spectrum shown on Figure~\ref{fig:spectra}, a Gaussian was successfully fit. From the standard deviation of the fitted Gaussian $A_2$, we can infer the observed velocity dispersion, $\sigma_{\mathrm{obs}}$ across the galaxy. $\sigma_{\mathrm{obs}}$ includes contributions from all observed dispersions, including beam smearing as well as the actual dispersion of the gas. The spatial distribution of $\sigma_{\mathrm{obs}}$ is shown in the right-hand panel of Figure~\ref{fig:SD}. The corresponding full-width half-maximum (FWHM) of each spectrum is $\mathrm{FWHM}_{\mathrm{obs}} = 2\sqrt{2\ln{2}}~\sigma_{\mathrm{obs}}$. 
We colour each spectrum in Figure~\ref{fig:spectra} based on the difference between the velocity of the Gaussian peak and the galaxy's average velocity. Red spectra therefore indicate areas that are red-shifted and blue spectra regions that are blue-shifted relative to the average velocity. We see there is a systematic red-to-blue gradient running across the galaxy, indicating the rotation of the galaxy. Because the gradient is smooth, we approximate the extent of the rotation of the galaxy via the velocity gradient $v_\mathrm{grad}$, which describes how much the molecular gas velocities between adjacent pixels differ. This observed gradient is present in $\sigma_{\mathrm{obs}}$, and must be corrected for in order to isolate the turbulent (unordered) motions (see \citet{Federrath:2016}). To determine $v_\mathrm{grad}$ in each pixel, we utilise the CARMA CO(1--0) observation's average velocity (1st moment) map. We estimate the local velocity gradient for a given pixel at coordinates $(i,j)$ as the vector sum magnitude of the difference in velocity between adjacent pixels, as in \citet{Varidel:2016aa}: 
\begin{align}
& v_\mathrm{grad}(i,j) = \\ \nonumber
& \quad\sqrt{[v(i+1,j)-v(i-1,j)]^2 + [v(i,j+1)-v(i,j-1)]^2}. 
\end{align}
(See also \citet{Zhou:2017aa} and \citet{Federrath2017SAMI}.) If a pixel has a neighbour that is undefined, the gradient in that direction is disregarded. 
The velocity gradient contributes to $\sigma_{\mathrm{obs}}$ through beam smearing. Beam smearing is caused when the velocity field changes on spatial scales smaller than the spatial resolution of the observation. The line-of-sight velocity dispersion is enhanced by beam smearing because rotational velocities can mimic dispersion. This leads to an artificial increase in the measured velocity dispersion if there is a steep velocity gradient across neighbouring pixels. We thus only regard pixels in which the observed velocity dispersion is at least twice that of the velocity gradient ($\sigma_{\mathrm{obs}}>2v_{\mathrm{grad}}$), resulting in pixels where beam smearing is not the dominant contribution to $\sigma_{\mathrm{obs}}$ \citep{Varidel:2016aa,Federrath2017SAMI,Zhou:2017aa}.
To obtain the final rotation-corrected velocity dispersion values across the galaxy, we subtract the velocity gradient from the observed FWHM in quadrature from every pixel:
\begin{align}
\mathrm{FWHM}_{\mathrm{corrected}} &= \sqrt{\mathrm{FWHM}_{\mathrm{obs}}^2 -  v_\mathrm{grad}^2} \\
\sigma_{\mathrm{corrected}} &= \frac{\mathrm{FWHM}_{\mathrm{corrected}}}{2\sqrt{2\ln{2}}}.
\end{align}
The beam-smearing corrected 2nd moment maps are shown in the right-hand panel of Figure~\ref{fig:SD}. The velocity dispersion is not isotropic across this galaxy, indicating that gravitational torques do not seem to act globally in this system. In particular, we find that the velocity dispersion is enhanced in the areas of the galaxy that are directly behind the direction of motion of the galaxy's bar. This indicates that the bar is likely driving turbulent motions. In the following we utilise these inferred velocity dispersions to explore the validity of a turbulence-regulated SF relation \citep{SalimEtAl2015} and compare them to $\sigsfr$ predicted by SF models that only account for gravity \citep{Kennicutt:1998aa,KrumholzDekelMcKee2012}.

\section{Comparing observed to predicted Star formation rates}
\label{sec:sfr_laws}

To determine whether turbulence influences the SF efficiency within NGC7674, we use the $\siggas$ and inferred velocity dispersion derived in Section~\ref{sec:veldisp} as inputs to predict $\sigsfr$ via the K98, KDM12 and \citet{SalimEtAl2015} (hereafter SFK15) SF relations. We compare these predictions to the observed $\sigsfr$ from the deep H$\alpha$ imaging to determine which, if any, of these models can best describe the small-scale physics taking place in NGC7674.
\subsection{Predictions from previous literature}

\begin{figure*}
\includegraphics[width=\linewidth]{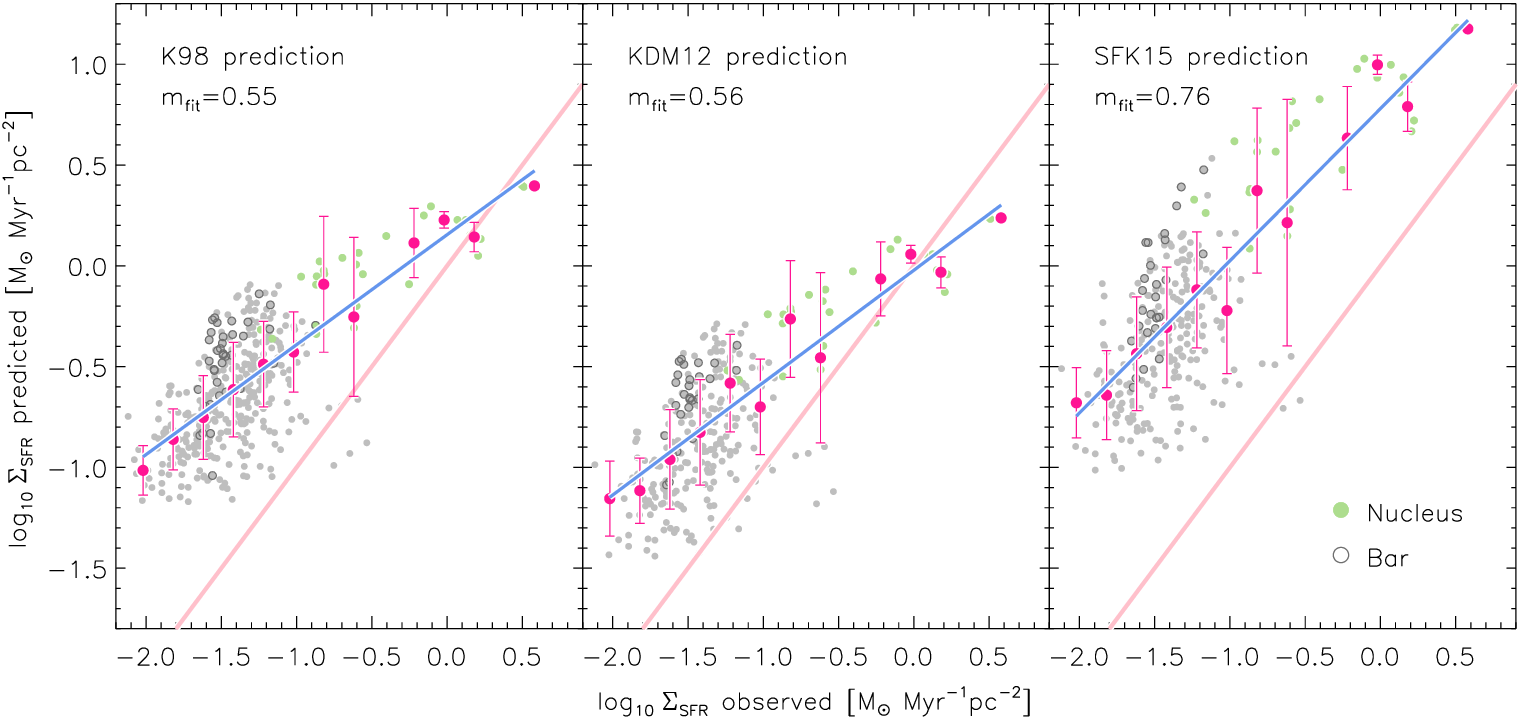}
\caption{Comparison between $\sigsfr$ predicted using SF relations of K98 (\textit{LEFT}), KDM12 (\textit{MIDDLE}) and SFK15 (\textit{RIGHT}) to that inferred from deep H$\alpha$ imaging. The overlay dark pink points are the average predicted $\sigsfr$ for 0.2 dex bins, with errors bars that show the standard deviation of each bin. The blue solid line is the best least-squares linear fit to the median predicted $\sigsfr$ values within each observed $\sigsfr$ bin. The pink solid line is the the one-to-one correlation (when the model fits the observations perfectly). The resultant slopes and offsets between the predicted and observed $\sigsfr$ for each of the K98, KDM12 and SFK15 SF models are tabulated in Table~\ref{tab:best_fits}. }
\label{fig:sfr_sfr_OLD}
\end{figure*}

This investigation has shown that areas of NGC7674 behave quite differently, and are washed out when only the global values are used. It is thus crucial to understand how small-scale interactions effect galaxy evolution, rather than relying on a global average. 

Descriptions of the K98 and KDM12 SF models were presented in Section~\ref{sec:intro}. In SFK15, the SF relation is derived accounting for the sonic Mach number of turbulence. The Mach number $\mach$ (i.e. the ratio between the absolute velocity of the gas and the local sound speed, $c_s$) is: 
\begin{align}
\mach = \sigma_{\mathrm{corrected}}/c_s.
\end{align}
In this study we assume a constant sound speed across the galaxy. To estimate the most viable value, we consider that all $\mathrm{H_2}$ gas is most likely to lie within the temperature range of 10 to 100 K, outside of which the gas will no longer be molecular under typical ISM conditions \citep{Ferriere:2001aa}. This range is supported by observations. Molecular clouds in Galactic spiral arms have been observed to have temperatures between $\sim 10-50$ K, whereas those in the Galactic centre can reach temperatures up to 100 K \citep{Ginsburg:2016aa}. The local sound speed of gas is related to the temperature by:
\begin{align}
c_s = \Big(\frac{k_BT}{\mu_\mathrm{p}m_\mathrm{H}}\Big)^{1/2},
\end{align}
where $k_B$ is the Boltzmann constant, $m_\mathrm{H}$ is the mass of a hydrogen atom and $\mu_\mathrm{p}$ is the mean particle weight. Under standard cosmic abundances, typical values of $\mu_\mathrm{p}$ are 2.3 for molecular gas and 0.6 for ionised gas \citep{Kauffmann:2008aa}. Thus we have that temperatures of $T=10$~K and $T=100$~K correspond to molecular sounds speeds of 0.2 km~s$^{-1}$ and 0.6 km~s$^{-1}$ respectively. In our calculation of Mach numbers and subsequent analyses we therefore choose a sound speed of $c_s = 0.4\pm 0.2$ km~s$^{-1}$. This estimate is appropriate for dense, cold star-forming phases of the ISM of temperatures of $T=10-100$~K. 

The SFK15 relation uses the fact that $\siggas$ and the density $\rho$ closely follow a log-normal probability density function (PDF) in both simulations and observations \citep{ KrumholzDekelMcKee2012, Federrath2013}:
\begin{align}
p(s)ds = \frac{1}{\sqrt{2 \pi \sigma_s^2}} \exp \Big(-\frac{(s - s_0)^2}{2\sigma_s^2} \Big) ds,
\end{align}
where the density variable $s$ is expressed as the logarithm of the density normalized to the mean density: 
\begin{align}
s = \ln(\rho/\rho_0)
\end{align}
and the logarithmic density variance $\sigma_s$ is defined as derived by \citet{PadoanNordlund2011} and \citet{MolinaEtAl2012}:
\begin{align}
\sigma_s^2 = \ln \Big(1 + b^2\mathcal{M}^2 \frac{\beta}{\beta +1} \Big).
\end{align}
$\sigma_s^2$ is itself parameterised by the turbulence driving parameter $b$, which represents the degree of compressive or solenoidal turbulence in the system \citep{FederrathGloverKlessenSchmidt2008, FederrathBanerjeeClarkKlessen2010}, the sonic Mach number $\mathcal{M}$, and the ratio between thermal and magnetic pressure $\beta$. SFK15 then derived their SF correlator by integrating over all densities for the entire PDF. The new correlator is denoted as the \textit{maximum} or \textit{ multi-freefall gas consumption rate}, $\gastmff$:
\begin{align}
\Big( \frac{\siggas}{t} \Big)_{\substack{\mathrm{multi-ff} \\ \mathrm{SFK15}}}  =  \frac{\siggas(\rho_0)}{\tff(\rho_0)} \int_{- \infty}^{\infty} \exp \Big(\frac{3}{2}~s \Big) ~p(s) ~ds ~.
\label{eq:integral_sfk}
\end{align}
By computing the integral and calibrating to Milky Way (MW) clouds and the Small Magellanic Cloud (SMC), the SFK15 model is reduced to: 
\begin{align}
\sigsfr &= 0.45\% \times \Big( \frac{\siggas}{t} \Big)_{\substack{\mathrm{multi-ff} \\ \mathrm{SFK15}}} \\
        &= 0.45\% \times \frac{\siggas(\rho_0)}{\tff(\rho_0)}  \times \Big( 1+ b^2 \mach^2 \frac{\beta}{\beta +1} \Big)^{3 / 8} .
\end{align}

SFK15 had assumed a natural mixed turbulence scenario of $b=0.4$. However, because of the likely presence of the bar in NGC7674, we expect that the gas in this galaxy may be subject to very strong shear forces \citep{Federrath2010,FederrathKlessen2012,Federrath2016CMZbrick}. A solenoidal driving scenario of $b=0.3$ is therefore a more reasonable choice in case of NGC7674 and therefore the value we assume in making our SFK15 SF predictions in this study. As in SFK15, for simplicity we assume the absence of magnetic fields, leading to a limit of plasma $\beta \rightarrow \infty$. We are able to measure Mach numbers across the galaxy from our gradient-corrected 2nd-moment map shown in Figure~\ref{fig:SD}, which represent as accurately as possible the velocity dispersion of the molecular gas.

The results of the $\sigsfr$ predictions derived from the K98 and KDM12 SF relations, as well as that of SFK15, are directly compared to $\sigsfr$ measurements inferred by deep H$\alpha$ imaging in Figure~\ref{fig:sfr_sfr_OLD}. The best least-squares linear fit to the median predicted $\sigsfr$ values within each observed $\sigsfr$ bin is shown as the blue solid lines. We apply a robust linear fit to the median $\sigsfr$ values within each 0.2 dex $\siggas$ bin instead of taking every individual point into account in order to minimise the influence of outlier lines of sight. We see that none of the  previous SF relations accurately capture the observed $\sigsfr$ in NGC7674. While the data points lie closer to the one-to-one line when the K98 and KDM12 relations are applied, the best fit lines exhibit slopes that are quite shallow, with power-law slopes of 0.54 and 0.55 respectively (with 1 being the ideal). This indicates that the K98 and KDM12 relations are fair at predicting intermediate SF where gravity may be the primary driver of SF, but fail at doing so at extremely low or high $\sigsfr$ regimes, when other influences of SF cannot be ignored. This may also explain why A15 found that NGC7674's globally sits near these relations. While the total values for $\siggas$ and $\sigsfr$ may be driven by the gravity of the dense nucleus, this information is insufficient to predict and explain the inefficiencies of SF outside of the nuclear region. The SFK15 prediction, which takes into account the internal turbulence within the small-scale structure of the system, exhibits a a power-law slope of 0.75, which is closer to unity, but still significantly off. While the correlation between observed and predicted $\sigsfr$ is better in SFK15 compared to K98 and KDM12, the SFK relation overestimates $\sigsfr$ by almost an order of magnitude overall. In the following, we address this discrepancy, by extending the SFK15 relation to incorporate the virial parameter (ratio of kinetic to gravitational energy), which had previously been left out of the SFK relation. Here however, because of the strong tidal interactions experienced by this HCG galaxy, we need to take the virial ratio of NGC7674 into account, as explained in the following.

\subsection{Extension to turbulence-regulated star formation relation: considering only gas denser than the critical density} \label{sec:theory-ext}
\begin{figure*}
\includegraphics[width=\linewidth]{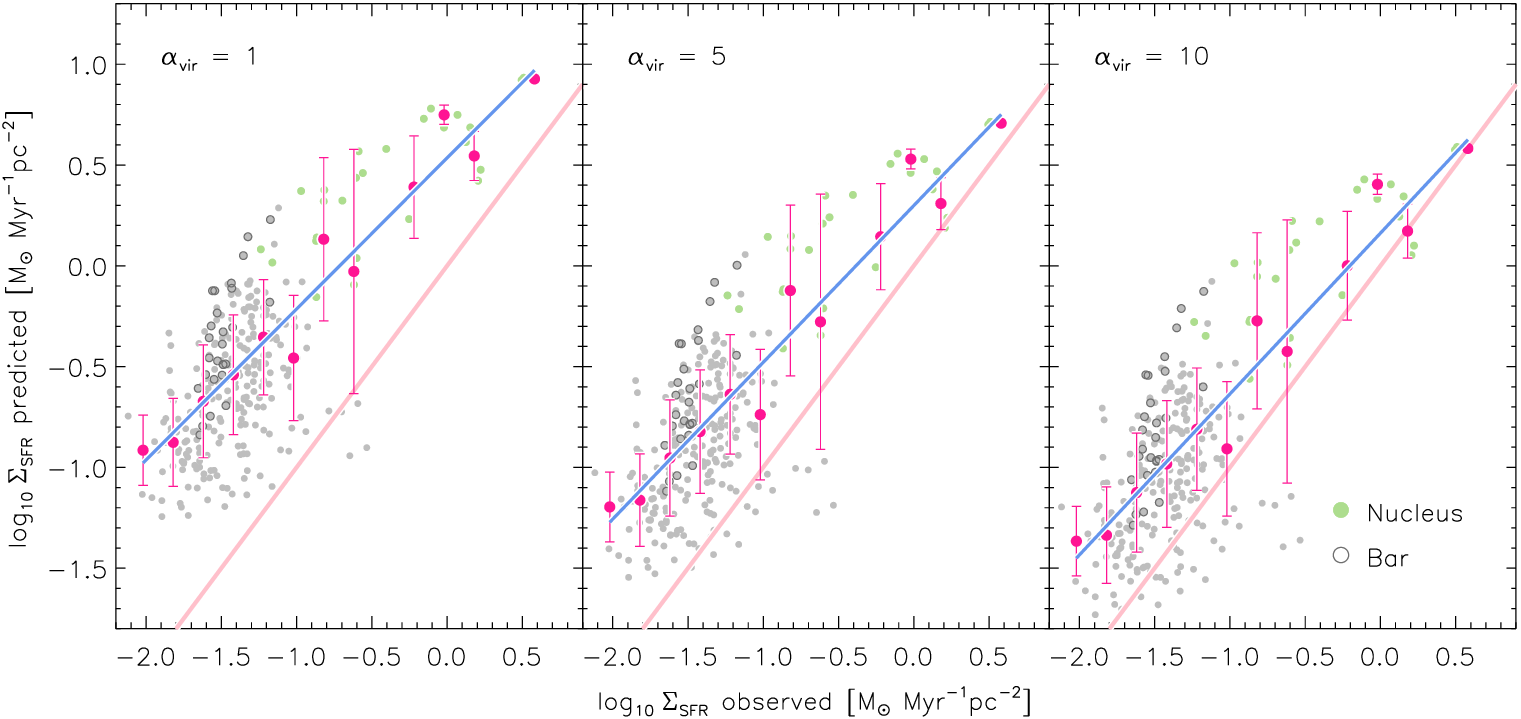}
\caption{Same as the right-hand panel of Fig.~\ref{fig:sfr_sfr_OLD}, but for the extended comparison between $\sigsfr$ predicted using this work's extended turbulence-regulated SF model (Equation~\ref{eq:final_new_law}) for assumed $\alpha_{\mathrm{vir}}$ of 1, 5 and 10 to the observed $\sigsfr$ inferred from deep H$\alpha$ imaging. The resultant slopes and offsets between the predicted and observed $\sigsfr$ for each of the $\alpha_{\mathrm{vir}}$ = 1, 5 and 10 scenarios for this SF model are tabulated in Table~\ref{tab:best_fits}.  }
\label{fig:sfr_sfr}
\end{figure*}

\begin{table}
\begin{center}
\begin{tabular}{ l c c}
SF prediction.                               &Offset & Slope \\
\hline
K98                                            &0.15    &  0.55   \\
KDM12                                          &-0.02    &  0.56    \\
SFK15                                          & 0.78    &  0.76    \\
This work, $\alpha_{\mathrm{vir}}=1$           & 0.54    &   0.75   \\
This work, $\alpha_{\mathrm{vir}}=5$           & 0.30   &   0.78   \\
This work, $\alpha_{\mathrm{vir}}=10$          & 0.16    &   0.80  
\end{tabular}
\end{center}
\caption{The offset and slope of the best least-squares linear fit to the median predicted $\sigsfr$ values within each observed $\sigsfr$ bin for the three previously derived SF models by K98, KDM12 and SFK15, as well as the extended turbulence-regulated SF relation derived in this work, for assumed $\alpha_{\mathrm{vir}}$ of 1, 5 and 10. A perfect correlation between predicted and observed $\sigsfr$ would result in an offset of 0 and a slope of 1.}
\label{tab:best_fits}
\end{table}

The SFK15 SF relation had integrated over the entire density PDF from $-\infty$ to $\infty$, suggesting that clouds of any density have the capability of forming stars. This assumption may be the cause of the overestimated $\sigsfr$ predictions in NGC7674. We explore a possible remedy by only considering gas dense enough to form stars. To test this, we integrate only from the minimum threshold density at which gas is able to collapse into stars. This minimum density is called the critical density, $s_{\mathrm{crit}}$, defined by: 
\begin{align}
s_{\mathrm{crit}} = \ln \Big[ \Big(\frac{\pi^2}{5}\Big) ~\phi_x^2 ~\alpha_{\mathrm{vir}} ~\mach^2 \Big] ~,
\end{align}
where $\phi_x$ is a fixed 'fudge factor' of order unity (introduced in \citet{KrumholzMcKee2005}; see \citet{FederrathKlessen2012} for details) and $\alpha_{\mathrm{vir}}$ is the virial parameter, which is the ratio of twice the kinetic energy to the gravitational energy, $\alpha_{\mathrm{vir}} = 2E_{\mathrm{kin}}/|E_{\mathrm{grav}}|$.

This $s_{\mathrm{crit}}$ was first defined in \citet{KrumholzMcKee2005} and is based on comparing the Jeans length to the sonic length. \citet{FederrathKlessen2012} extended this derivation by including magnetic fields in the description, finding $s_{\mathrm{crit}}$ by comparing the magnetic Jeans length with the magneto-sonic scale. (For the definition and relevance to the sonic length see \citet{FederrathKlessen2012} and \citet{Federrath:2016}). We assume the \citet{FederrathKlessen2012} fudge factor value of $\phi_x = 0.17 \pm 0.02$. This value was determined by fitting to a set of 34 numerical simulations of molecular clouds with resulting SF rates over more than two orders of magnitude. To incorporate this into our model we take the $\gastmff$ parameterisation of SFK15 shown in Equation~\ref{eq:integral_sfk} and integrate from $s_{\mathrm{crit}}$ instead of $-\infty$: 

\begin{align}
\Big( \frac{\siggas}{t} \Big)_{\substack{\mathrm{multi-ff} \\ \mathrm{EXTENDED}}}  &=  \frac{\siggas(\rho_0)}{\tff(\rho_0)} \int_{s_{\mathrm{crit}}}^{\infty} \exp \Big(\frac{3}{2}~s \Big) ~p(s) ~ds  \\
                                                                                   &= \frac{\siggas(\rho_0)}{\tff(\rho_0)} \exp \Big(\frac{3}{8}~\sigma_s^2 \Big)\frac{1}{2} \Big[ 1+ \mathrm{erf}\Big(\frac{\sigma_s^2 - s_{\mathrm{crit}}}{\sqrt{2 \sigma_s^2}} \Big)\Big] \\
                                                                                   &=  \frac{1}{2} \Big[ 1+ \mathrm{erf}\Big(\frac{\sigma_s^2 - s_{\mathrm{crit}}}{\sqrt{2 \sigma_s^2}} \Big)\Big] \Big( \frac{\siggas}{t} \Big)_{\substack{\mathrm{multi-ff} \\ \mathrm{SFK15}}} .
\end{align}

Assuming the same SF efficiency as in SFK15, we derive an extended SF relation which is more physically descriptive,
\begin{align}
\sigsfr &=  0.45\% \times \Big( \frac{\siggas}{t} \Big)_{\substack{\mathrm{multi-ff} \\ \mathrm{EXTENDED}}} \\
&= 0.45\% \times \frac{1}{2} \Big[ 1+ \mathrm{erf}\Big(\frac{\sigma_s^2 - s_{\mathrm{crit}}}{\sqrt{2 \sigma_s^2}} \Big)\Big] \Big( \frac{\siggas}{t} \Big)_{\substack{\mathrm{multi-ff} \\ \mathrm{SFK15}}}.
\label{eq:final_new_law}
\end{align} 

Note that for the limiting case of $\alpha_{\mathrm{vir}}$=0, the extended SF relation reduces exactly to the previous SFK15 relation (Eq. 18). In practice, the predicted $\sigsfr$ for systems with $\alpha_{\mathrm{vir}} < \approx 1$ is nearly identical to the previous SFK15 relation. Thus, the new relation only makes a significant difference for systems with $\alpha_{\mathrm{vir}} > 1$.

Similar to the method in Section~\ref{sec:sfr_laws} we now use the CO(1--0) observations of $\siggas$ and inferred velocity dispersion derived in Section~\ref{sec:veldisp} as inputs in our newly derived extended turbulence regulated SF relation. We compare these predicted values to the $\sigsfr$ from deep H$\alpha$ imaging to investigate whether this new SF relation better describes the internal physics of NCG7674. 



\subsection{Applying extended turbulence-regulated star formation relation to NGC7674}
We apply the extended turbulence-regulated SF relation derived in Section~\ref{sec:theory-ext} to predict $\sigsfr$ across NGC7674 and investigate how these compare to observations. In the absence of of a direct measurement of $\alpha_{\mathrm{vir}}$, we assume fixed, uniform values of $\alpha_{\mathrm{vir}}$ across the galaxy in order to investigate how $\alpha_{\mathrm{vir}}$ impacts $\sigsfr$. Figure~\ref{fig:sfr_sfr} shows how these predictions compare to observations for assumed $\alpha_{\mathrm{vir}}$ values of 1, 5 and 10. The $\alpha_{\mathrm{vir}}$ values that best predict $\sigsfr$ in NGC7674 are $\alpha_{\mathrm{vir}}$ $\geq$10.

For the same values of $\siggas$ and $\mathcal{M}$, Equation~\ref{eq:final_new_law} results in the SFK15 SF prediction multiplied by a factor dependent on $\alpha_{\mathrm{vir}}$. Since the critical density $s_{\mathrm{crit}}$ is directly proportional to $\alpha_{\mathrm{vir}}$, systems with higher values of $\alpha_{\mathrm{vir}}$ would in turn have higher values of $s_{\mathrm{crit}}$ and consequentially the inclusion of a smaller fraction of the density PDF capable of forming stars. This ultimately results in a prediction of more suppressed SF at higher values of  $\alpha_{\mathrm{vir}}$. Using Equation~\ref{eq:final_new_law}, predicted $\sigsfr$ get closer to the observed $\sigsfr$, with increasing $\alpha_{\mathrm{vir}}$. This is reflected in Figure~\ref{fig:sfr_sfr}, where we see a decrease in offset of the best robust linear fit to the data in the logarithmic $\sigsfr~(\mathrm{observed}) - \sigsfr~(\mathrm{predicted})$ plane with increasing $\alpha_{\mathrm{vir}}$, but a nearly constant slope. The resultant slopes and offsets between the predicted and observed $sigsfr$ for each of the  SF models tested in this study are tabulated in Table~\ref{tab:best_fits}.

\section{Discussion} \label{sec:discussion}

It is plausible that this system possesses sufficient excess kinetic energy that $\alpha_{\mathrm{vir}}$ of 10 or more is viable. The HCG environment of NGC7674 exhibits $\mathrm{H_2}$ shocks \citep{Cluver:2013aa}, indicative of excessive kinetic energy. Environments with conditions thought to be similar to that of Hickson compact groups, such as the Central Molecular Zone (CMZ) of the Milky Way, have also been found to exhibit higher virial parameters \citep{Kruijssen:2015aa, Federrath2016CMZbrick}. The CMZ has also been observed to have a low efficiency of SF.

\citet{Federrath2016CMZbrick} found that the virial parameter within the MW cloud nicknamed "the Brick" exhibits a value of about 4. The fact that the objects in our study are experiencing gravitational torque via their companion galaxies makes it possible that these environments could exhibit more extreme $\alpha_{\mathrm{vir}}$ than the clouds in the CMZ. It is thus reasonable to believe that if a MW cloud can exhibit a virial parameter of 4, these effects may be amplified in a HCG environment to induce high virial parameters of 10 or more.

Furthermore, there is strong evidence that the turbulence in this system is induced by the presence of a bar. We see in the spatial distribution of inferred velocity dispersions shown in the right-most panel of Figure~\ref{fig:SD} that the top left and bottom right quadrants of the centre region of the galaxy exhibit the highest velocity dispersions. Just as a solid moving through a fluid will create vortices behind it, the rotation of the galaxy's bar through the ISM will disturb the gas which it has already passed, inducing heightened turbulent motions just behind the bar. Another competing factor that may be enhancing the velocities of molecular gas in this system is the torquing of gas into the centre bulge due to the existing concentration of mass and therefore gravitational forces \citep{ForbesKrumholzBurkert2012Gravity, Forbes2014Gravity, Krumholz:2018aa}. However, if gravity was the primary driver of motion in this system, the spatial distribution of regions of high velocity dispersions would look more isotropic across all four quadrants of the galaxy plane. This further suggests that disruption from the bar and therefore strong shear and solenoidal driving scenarios are significant sources of kinetic motions in this galaxy. These enhanced kinematic properties, coupled with strong torques from neighbouring galaxies, makes it highly likely that high $\alpha_{\mathrm{vir}}$ of 10 or more is plausible within this system. 

\subsection{Limitations and caveats}
In this study we choose to explore how the SF predictions using the extended turbulence-regulated SF relation (Equation~\ref{eq:final_new_law}) behaves with increase of $\alpha_{\mathrm{vir}}$ uniformly across the galaxy. We choose this method instead of individually estimating a value of $\alpha_{\mathrm{vir}}$ because we do not have sufficient resolution to accurately measure this quantity, as this would require a full energetics analysis of this system, which is beyond the scope of this study. 


Moreover, we acknowledge that the resolution of the data used in this study has been too coarse to resolve the GMC scales needed to isolate individual regions of distinct kinematic properties. In this study we compromised high resolutions for reliability of the parameters we input into Equation~\ref{eq:final_new_law}, but subsequent studies should strive at least for GMC scaled resolutions to scrutinise the legitimacy of this description of SF. For this we would need data which could resolve at least GMCs, which scale around a few tens of parsecs each \citep{Krumholz2017SFtextbook}. 


Finally, we emphasise that the simplistic assumption that magnetic fields are absent in this system is definitely unrealistic. Magnetic fields have been observed in almost all extragalactic sources \citep{FederrathKlessen2012, Krumholz2017SFtextbook}. In the presence of magnetic fields, the magnetic pressure $P_{\mathrm{mag}}$ is greater than the thermal pressure $P_{\mathrm{th}}$. Thus the plasma $\beta$, which is a ratio of these two values $\beta = P_{\mathrm{th}}/P_{\mathrm{mag}}$ will be small in the presence of strong magnetic fields. In both the SFK15 SF model and this work's Equation~\ref{eq:final_new_law}, this will result in a reduction in the amount of $\sigsfr$ predicted to be in any system when compared to assuming complete lack of magnetic fields. Numerical simulations have also supported the presence of star formation suppression as an effect of introducing magnetic fields \citep{PadoanNordlund2011,FederrathKlessen2012,Federrath:2015aa,Krumholz:2019aa}.

\section{Conclusions} \label{sec:conc}
In this study we present and compare $\siggas$ inferred from CO(1--0) and $\sigsfr$ inferred from deep H$\alpha$ imaging of the face-on Hickson Compact Group galaxy NGC7674. We measure velocity dispersions across the galaxies via the CO(1--0) channel maps and piece together this information to explore the effects and roles of turbulence in shaping the kinematics and star formation inefficiency of this galaxy. In doing so we find the following: 
\begin{itemize}
\item NGC7674 does not exhibit a straightforward relation between $\siggas$ and $\sigsfr$, with any value of $\siggas$ correlating to a $\sigsfr$ range of up to an entire order of magnitude on the KS plane. We find that this galaxy exhibits two distinct SF sequences in the KS plane; one for the nucleus of the galaxy and one for outlying regions. 
\item We find that spatially, the densest regions of $\siggas$ do not necessarily correspond to regions of the shortest depletion times; instead the exact opposite is often the case in many sections of this galaxy outside the nucleus.
\item The velocity dispersion across NGC7674 is non-isotropic and enhanced in areas directly behind the direction of motion of where we believe the galaxy's bar lies. This suggests that the bar is key physical feature driving turbulence within this galaxy. The interplay between turbulence induced by the bar, accretion of gas into the galaxy's dense nucleus and competing forces from the torques induced by neighbouring galaxies in the group all contribute to the likelihood that this galaxy experiences extreme, enhanced kinematic properties. 
\item We apply the K98, KDM12 and SFK15 SF relations to predict $\sigsfr$ and compare them to observations from deep H$\alpha$ imaging. We find that none of these relations adeptly predict the observed $\sigsfr$. Whilst the global values of NGC7674 sit near the K98 and KDM12 SF relations, when isolating individual lines of sights it is clear that these models do not result in a linear correlation between the predicted and the observed $\sigsfr$. The SFK15 relation represents a more linear correlation between the model and observed $\sigsfr$ than that predicted by K98 and KDM12. However, the SFK15 model systematically overestimates the $\sigsfr$ by almost an order of magnitude on this galaxy. We attribute this overestimation to SFK15's assumption that gases of all densities have the potential to form stars. 
\item We extend upon the theoretical SF relation by SFK15 to formulate a new SF relation, Equation~\ref{eq:final_new_law}. We do so by integrating the density PDF from the critical density $s_{\mathrm{crit}}$ to formulate an SF relation dependent on $\siggas$, $\mathcal{M}$ and the virial parameter $\alpha_{\mathrm{vir}}$. We did this in response to the finding that the previously formulated SF relations by K98, KDM12 and SFK15 did not adequately describe the small-scale physics of this system. 
\item We find that Equation~\ref{eq:final_new_law} describes the SF across NGC7674 best in the assumption that the galaxy exhibits high $\alpha_{\mathrm{vir}}$ values of $\geq$10. Such high values of $\alpha_{\mathrm{vir}}$ are plausible in NGC7674 due to the presence of a bar which is likely to be inducing turbulence behind its direction of motion, as well as the external torques being exerted by the neighbouring galaxies in the Hickson compact group. It is very likely that these physical features are injecting excess kinetic energy into this system, which may cause high values of $\alpha_{\mathrm{vir}}$. 
\end{itemize}

  We anticipate that the extension of SFK15's turbulence-regulated SF model to account for suppression of SF in systems with high excess of kinetic energy will be utilised to explore potential reasons why inefficient systems such as galaxy groups, galaxy clusters or post-starburst systems exhibit so little SF for the excess of gas present in the system. We hope that future observations and techniques will directly measure $\siggas$, $\sigsfr$ and kinematic components in Galactic and extragalactic sources to at least the $\approx10$ parsec GMC-scale resolutions, if not smaller. We hope future studies will take advantage of such observations to test whether Equation~\ref{eq:final_new_law} accurately describes the physics of SF on small scales, and therefore evaluate the validity of a multi-freefall description of SF.  

\acknowledgments
Parts of this research were supported by the Australian Research Council Centre of Excellence for All Sky Astrophysics in 3 Dimensions (ASTRO 3D), through project number CE170100013. C.~F.~acknowledges funding provided by the Australian Research Council (Discovery Project DP170100603 and Future Fellowship FT180100495), and the Australia-Germany Joint Research Cooperation Scheme (UA-DAAD).

\bibliographystyle{aasjournal}
\bibliography{HCG_TURBULENCE}

\end{document}